# Artificial Intelligence and Innovation to Reduce the Impact of Extreme Weather Events on Sustainable Production

Derrick Effah, Chunguang Bai, and Matthew Quayson

**Abstract**
Frequent occurrences of extreme weather events substantially impact the lives of the less privileged in our societies, particularly in agriculture-inclined economies. The unpredictability of extreme fires, floods, drought, cyclones, and others endangers sustainable production and life on land (SDG goal 15), which translates into food insecurity and poorer populations. Fortunately, modern technologies such as Artificial Intelligent (AI), the Internet of Things (IoT), blockchain, 3D printing, and virtual and augmented reality (VR and AR) are promising to reduce the risk and impact of extreme weather in our societies. However, research directions on how these technologies could help reduce the impact of extreme weather are unclear. This makes it challenging to emploring digital technologies within the spheres of extreme weather. In this paper, we employed the Delphi Best Worst method and Machine learning approaches to identify and assess the push factors of technology. The BWM evaluation revealed that predictive nature was AI's most important criterion and role, while the mass-market potential was the less important criterion. Based on this outcome, we tested the predictive ability of machine elarning on a publilcy available dataset to affrm the predictive rols of AI. We presented the managerial and methodological implications of the study, which are crucial for research and practice. The methodology utilized in this study could aid decision-makers in devising strategies and interventions to safeguard sustainable production. This will also facilitate allocating scarce resources and investment in improving AI techniques to reduce the adverse impacts of extreme events. Correspondingly, we put forward the limitations of this, which necessitate future research.

Keywords; Extreme Weather, Sustainable Production, Technology, and Innovation.

## 1.0 INTRODUCTION

The occurrence of extreme weather has been predominant in recent years(Weilnhammer et al., 2021). For example, 2019/2020 saw unprecedented bushfires in southeast Australia. The vast forest burnt area, loss of wildlife, the fire's radiative strength, and an unusual number of fires that escalated into extreme events were beyond historical records(Abram et al., 2021). The impact of extreme events on the loss of life is considerably higher. This could translate into reduced human resources in production and operation, which is catastrophic for supply and production chains. In 2020, the UK was the third warmest, fifth wettest, and eighth sunniest in historical climate records(Chik & Xue, 2021). In China, in mid-July 2021, the city of Zhengzhou was submerged by floodwater. The city received a year worth of rain in just three days(Chik & Xue, 2021). 2010 extreme winter conditions disrupted freight and railways, coincidentally affecting logistics and business operations(Ludvigsen & Klæboe, 2014). Harsh weather conditions result in taunt shortages of resources for production and operations into a considerable economic downturn. The health risk of workers and labor shortages in post-extreme events are detrimental to sustainable production and supply resilience.

Agriculture production is not spared in during extreme events. Between 2015 and 2016, South Africa recorded a drought (Vetter et al., 2020), which caused a decline in livestock production. About 43% of herds were lost due to extreme weather events. Authors (Vetter et al., 2020) reported that cattle numbers remained stunt for three years after the drought. Across the globe, Africa is seen as the most vulnerable to climate risk and disasters since the region is exposed to high climate extremes and low adaptive capacity(Awojobi & Tetteh, 2017). For instance, frequent cyclones in Madagascar destroyed food crops and infrastructures, leaving many homeless and Militating against achieving the UN sustainable development goals, especially goals 1(no poverty), 2 (Zero hunger), and 15 (life on earth). Food security is an undisputable subject of debate; however, most agriculture systems in Africa are rain-fed due to less mechanized farming.

Moreover, the slightest unpredictable, harsh weather conditions will jeopardize the livelihood of many.

Extreme events have very low potential to occur under normal conditions unless there is a disruption in the natural system, for example, human-induced biodiversity loss and climate change. These are particularly unpredictable and with a drastic change in a habitable system. Farmers in a classical emerging country lamented over poor crop yield and severe fires due to unpredictable extreme weather (Asante et al., 2017), making accurate, reliable, and accessible weather prediction crucial to sustainable agriculture and the design of early warning response.

Predictable weather is imperative to inform farmers' decisions on which climate-resilient crop to plant at different times throughout the year. In the context of weather predictions, real radar data, UNet, and advection approaches have been used. Although these methods are based on robust numerical weather forecasts obtained by solving mathematical equations, they struggle to find fine-grained forecasts. Building a generalized module with less precision will result in inaccurate predictions. Fortunately, AI has gained success in pattern recognition, health, marketing, and business, with high precision and accuracy. As such, it could be harnessed in predicting weather and improving one early warning response. For instance, an accurate prediction of recent Australian fires could have saved thousands of lives, if not millions, including animals and humans. The impacts of the wildfires would have been reduced, and enterprises could have built concrete and strategic plans for handling business post-fires. For instance, the Australian wildlife department could have found a safe ground for most wild animals, reduced the forest fuel, relocated the forest-fringed community, and moisturized high-risk wildlands to reduce the impact of fires.

Forecasting and designing real-time weather maps, disseminating information, and instant printing tools or materials will contribute to early warning, suggesting resilient crops, planting times, and routing systems regarding the weather and climate conditions. The characteristics of AI make it a better choice for reducing the impact of extreme weather. However, some questions still need clarity from the expert's perspective. These questions made up our research questions.

1. How will AI achieve the goal of reducing the adverse impacts of extreme events?
2. What are the attributes of AI which will make it possible?
3. What are the technological platforms required by AI?
4. What are the potential challenges of adopting AI?

Positive outcomes are achievable with the transformation of modern cities into a carbon-neutral environment. In this paper, we contributed to addressing the issues of how artificial intelligence and supporting technologies could reduce the impact of extreme weather for sustainable production. We employed the unified theory of acceptance and use of technology (UTAUT) to categorize the attributes of AI using the Delphi method. The first phase of Delphi was to obtain AI's attributes, challenges, and supporting technologies. The CVI was used as a robustness check to validate the responses of the Delphi before utilizing the BWM methodology to evaluate the attributes and enablers of AI against the four dimensions of UTAUT (Performance expectancy, effort expectancy, social influence, and facilitating conditions). Finally, machine learning techniques were used to predict the forest fire burn area of a publicly available dataset as a test case to affirm the predictive potential of AI.

This study aims to stimulate scholarly discussions on the role of modern technologies and innovation in reducing the impact of extreme weather on sustainable production. It responds to the call for broader investigation (Choi et al., n.d.). We contribute by unearthing the critical issues of extreme weather and appropriate technological innovations to minimize the consequences. Also, we propose a conceptual framework and scenario for digital technologies to reduce extreme weather. Again, we propose further research directions that will guide further studies into technological innovations for reducing extreme weather effects on sustainable production, especially in emerging economies.

The paper progresses as follows; section 2 discusses the literature on extreme events and operation management, capabilities of AI on sustainable production and operations, and analysis methods for extreme weather events. Section 3 outlines the methodologies utilized in achieving the goal of this study and proposes a conceptual scenario of AI and supporting technologies' roles in reducing the adverse impact of extreme weather events. We presented the results and discussion in the section. The study's implication and conclusion were presented in sections 5 and 6, respectively.

## 2.0 Literature Review

### 2.1 Extreme Weather Events and Operation Management

Extreme weather events pose a significant threat to supply chains and operation management. As indicated by (Dwivedi et al., 2018), governments and NGOs operate and maintain global supply chains to assist vulnerable populations in meeting their needs. These buffered supply chains are designed to carry out and distribute food, water, sanitation, and medication throughout many regions across the globe (Shareef et al., 2019). In situations of extreme events, there are possible difficulties in operating and coordinating emergency supply chains due to poor management, inefficiencies, disruptions risk, and negative structural dynamics-related issues (Dwivedi et al., 2018). Unpredictable extreme events along supply chains are subjected to disruptions risk. Efficiency and cost-effectiveness are typically less crucial in emergency situations, whereas resilience, efficiency, and responsiveness are given more weight (Rameshwar Dubey et al., 2017). Delays in timely logistics, facility failure, and operability issues among supply chain entities (Rameshwar Dubey et al., 2017)(Dwivedi et al., 2018)are key impacts on sully chain and operations management. Developing contingency plans and options are required to boost supply chain resilience(Shareef et al., 2020). Here, AI and supporting technologies could play a huge forecasting and decision-making role in improving supply chain resilience.

The impact of extreme events on agrarian economies is detrimental to sustainable agricultural production, food security(SDG goal 2), and poverty reduction(SDG goal 1). These impacts are more predominant in vulnerable regions such as sub-Saharan Africa (Codjoe & Atiglo, 2020). A study by (Rameshwar Dubey et al., 2017) demonstrated how the provision of electricity services is interrelated to extreme weather events with frequent energy outrages with severe rains and floods in Ghana. Indirectly, electricity outrages hinder efficiency and productivity.

The 1983 drought in Ghana caused significant crop failures and prolonged famine(Kushitor, 2021). Ghana is an agrarian economy, and food production has met the menace of natural fires. Extreme events in current times threaten the food basket of Ghana and elsewhere, which needs timely technological and policy response to build a resilient agriculture and food supply chain. Plant breeding techniques to develop resistant crops to fires and extreme heat are genomic advancements that can leverage the power of AI technologies(Qaim, 2020). Unpescribe burning in agricultural landscapes substantially escalates to wildfires given the drivers(extreme weather) of fire occurrence. The economic cost of Indonesia fires exceeded US $16 billion in 2015(Edwards et al., 2020). Digitization of the agriculture system will give easy monitoring of farming practices to prevent unwanted fires and, consequently, sustainable agriculture and food production(Quayson et al., 2020)(Basso & Antle, 2020).

Across the globe, the economic loss from extreme events is tremendously higher, especially in floods. According to "our world in data" statistics, the cost of floods amounts to about 7.7 billion dollars. In retrospect, research on extreme events has focused on their effects on yield and food production, education(Groppo & Kraehnert, 2017), and budget balance (Kreft et al., 2016). Research on reducing the impact of extreme events has taken the pace of insurance (Kunreuther, 2015), risk analysis and economic incentive(Kunreuther et al., 2004), and adaptation strategies(McDaniels et al., 2008). These researches analyze the social dimension of extreme weather, which is comparatively relevant, but how emerging technologies will play a role in reducing the adverse impacts of extreme has not been explored in current research.

### 2.2 The Capabilities of AI for Sustainable Production and Operations

AI, which encompasses machine learning and deep learning, was designed to mimic human intelligence in performing a task and making decisions assuming human intelligence (Bughin et al., 2018)(Dick, 2019). With a growing need for businesses to be more competitive, adapt to global issues, and be more environmentally friendly, research on AI-enabled business operation are of high priority.

(Ransbotham et al., 2018) posits data as a key enabler for AI applications. In recent years organizations have captured huge datasets with varying data structures from multiple sources and formats(Kersting & Meyer, 2017)(Ghasemaghaei & Calic, 2019). The quality of these data is highly prioritized in designing AI with high decision-making, learning, and prediction capabilities. A classical challenge to an organization in advancing AI is data management in the form of labeling to build supervised AI applications. Organizations could harness internal and external data to provide a competitive environment to boost organizational sustainable production and operations (Mikalef & Gupta, 2021). There a barrier to full-scale transitioning into AI-enabled production and operations but entrepreneurial orientations(EO) asserts operational performance in an AI. The findings of (Rameshwar Dubey et al., 2020) confirm the ability of EO to push organizations to explore big data and AI-powered systems to achieve greater operational performance. This also means organizations could shape their dynamic capabilities to meet environment dynamism to improve organizational performance.

The capabilities of AI are enabled not only by data but also by technological infrastructure, human resources, and skill-set. (Chui & Malhotra, 2018) published a 2018 report indicating that technological infrastructure is one of the major barriers to AI adoption. Particularly, organizations with fewer resources find it cumbersome to compete in an AI environment (Dwivedi et al., 2021). For instance, deep learning systems must be updated data constantly since they can retrain themselves as they work. This foster the learning capabilities of AI and consequently improves the prediction and decision-making capabilities. Infrastructure investment is required to achieve the full capabilities of AI in an organization (Mikalef & Gupta, 2021). Leveraging AI capabilities in automating systems changes the phase of traditional manufacturing and operations(Kakani et al., 2020). (Kakani et al., 2020) stressed that the growth of automation has led to productivity peaks in the manufacturing sector and modern industry within a short period.

Also, in the wake of extreme events and unpredictable disasters, humanitarian supply chain management is highly important to limit the suffering of victims and promote rehabilitation. The findings of (R Dubey et al., 2022) indicated that AI-driven big data analytics capabilities are crucial in determining the resilience, agility, and performance of humanitarian supply chains. It is imperative to reduce the impacts of extreme events, but How can AI help in this context? The authors (Pournader et al., 2021) revealed in their review that supply chain operations still lag in adopting current technological trends. There is less research on the roles of AI and supporting technologies in reducing the impact of extreme weather for sustainable production and supply chain operations.

**2.3 Analysis method for Extreme Weather Events**

The adverse effect of extreme events spreads across all sectors of society. In the health sector, (Chique et al., 2021) used an exploratory approach to assess psychological impairment due to extreme weather. (Tian et al., 2022) used comparative analysis to reveal the severity of extremeness in China. (Rimi et al., 2019) employed a quantitative assessment of the risk of extreme events using large ensembles of regional climate modeling systems. The authors claimed this tool could be used to study extreme weather events. The author (Walsh & Patterson, 2022) utilized a Lowless model and a series of wavelet transforms to observe the periodicity in extreme weather events. The authors (Martinez-Pastor et al., 2021) utilized a mapping and sensitivity analysis technique to enhance the decision-making process for a resilient transport network.

Disruption in transportation taunts the distribution of goods and services and the entire supply chain. (Ali et al., 2021) modeled supply chain risks factoring in extreme events as the most common type of it. The authors employed a hybrid Delphi and fuzzy arithmetic hierarchy process(AHP) to ascertain supply chain risks. Research by (Cowled et al., 2022) revealed that about 465% of farms reported a high death rate of their livestock, more than 17 million hectares of land were burnt, and about 300 homes were destroyed. Beyond fires, the operational activities of Japanese businesses were also impeded by the earthquakes that struck Japan in 1995, 2007, and 2011. As a result, businesses lost their crucial competitive positions in international supply chains. However, the above methods focus on the risk and evaluate the occurrence of extreme events without giving a classical analysis of how modern technologies could be harnessed to reduce the impact of extreme events. For instance, the operational activities of Japanese businesses were also impeded by the earthquakes that struck Japan in 1995, 2007, and 2011. As a result, businesses lost their crucial competitive positions in international supply chains. The unpredictability of extreme events disturbs social systems and production chains, making predictive models technologies relevant.

Most research on weather prediction and forecast utilized real-time radar(Kouwen, 1988)(Seo & Krajewski, 2020), UNet (Trebing et al., 2021)(Fernández et al., 2022), and advection approaches(Inage, 2019) which have difficulties in obtaining fine-grained forecast. These models depend on solving physical equations. Nevertheless, these mathematical methods fail to be precise. Unlike these models, AI-based techniques can predict weather events with high accuracy and precision. Also, research on technological advancement in extreme weather is still few, with little focus on the roles of current technologies in reducing the impact of extreme events. Data availability and quality are compelling challenges to achieving high-performance AI models. The limited research on AI and extreme weather events have not tackled the data scarcity issues in training ML models. Thus the contribution of our paper is stated in section 1.

### 3.0 Methodology and Case Study

In this section, we develop an AI conceptual model for predicting and reducing the adverse impacts of extreme weather events using Delphi, BWM, and machine learning. First, Delphi is used to identify the various relevant and critical attributes of using AI in extreme weather prediction. Second, the Unified Theory of Adoption and Technology Use of Technology (UTAUT) is adopted to classify those critical attributes. Third, BWM is introduced to evaluate the weights of those critical attributes. Finally, based on the weights of those critical attributes, we develop a machine learning model to predict and reduce the impact of extreme weather events. Before developing the AI conceptual model, we first introduce the relevant background of this case.

### 3.1 Background of the field case

Extreme weather events affect all areas of society, including the environment, education, agriculture, production, transportation, and social economy. The death rate in Ghana from natural fires and burns is the highest globally, though most of the global natural fires occur outside Ghana (Figure 1). Cocoa farmers in Ghana complain about unpredictable natural fire and their impact on cocoa production (Asante et al., 2017). In December 2020, a forest fire destroyed about 300 acres of soybean and cowpea farms in the northern part of Ghana. Tzibaa farm, the victim of the forest fire, invested nearly ¢700,000 (Ghana currency, which represents $93,000), and its investment failed to produce the expected return (Joy online, 2020). These negative impacts on a cocoa-producing country will affect the entire supply and food value chain, which is not conducive to food security and poverty reduction. AI and other digital technologies are promising to reduce the negative impact of extreme weather events. The researches focus on assessing the role of AI in reducing the impact of extreme weather events, which makes Ghana a laudable case of study. However, in the context of Ghana, the roles of AI in reducing the impact of extreme weather events are less explored. Although the field case is located in Ghana, the results of this study can also be applied to other places.

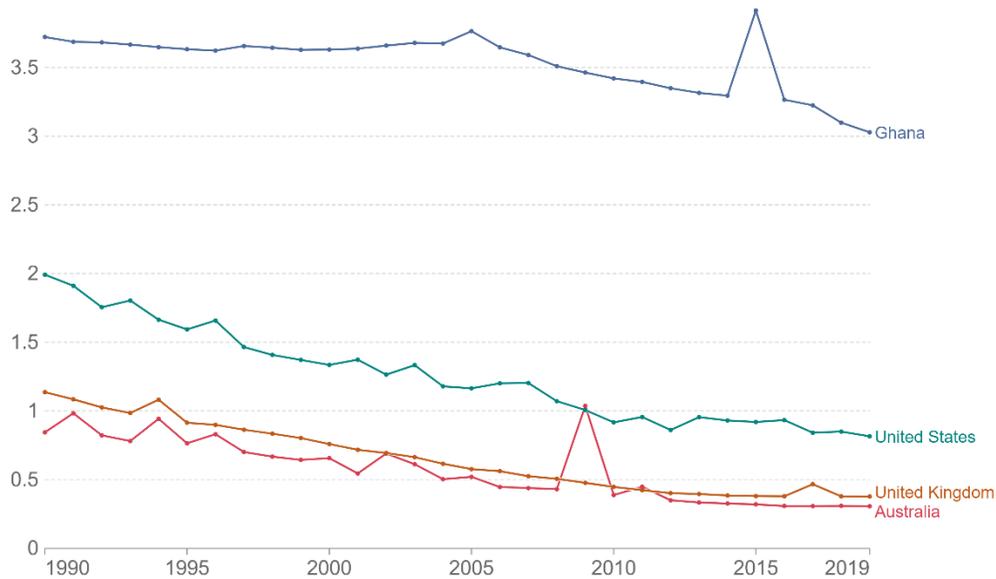

Figure 1 Deaths rate due to fires and burns

### 3.2 Delphi method

Delphi method, also known as Estimate-Talk-Estimate or ETE, which was introduced by Oracle of Delphi in 1946 (Linstone & Turoff, 2018)(Belton et al., 2019) , is essentially a feedback anonymous structured communication technique or method based on a panel of experts. Its general process is to sort out, summarize and count the problems to be predicted after obtaining experts' opinions, and then anonymously feed them back to various experts to solicit opinions again, then concentrate and feed them back until they reach a consensus.

In our study, the Delphi method was used to identify the various relevant and critical attributes of the use of AI in extreme weather prediction, as seen in Table 1. To do this field analysis, we recruited decision-makers who are experts on the subject matter through the invitation letter, which included the purpose of this study and its duration. First, 13 experts were identified as using the past experience approach, 9 experts were willing to invest time in the research process. All experts are the decision-makers in technology and disaster response management units, as seen in Table 1.

Table 1 Demographics of Decision-Makers and Experts

| Experts for Delphi | Educational level | Years of experience | Industry | Position |
|---|---|---|---|---|
| Expert 1 | Tertiary | 16 | Technology Company | Software engineer |
| Expert 2 | Tertiary | 27 | National Disaster Management Organisation | Research Scientist |
| Expert 3 | Tertiary | 22 | Technology Company | Sales Engineer |
| Expert 4 | Tertiary | 25 | Telecommunication company | Digital production manager |
| Expert 5 | Tertiary | 21 | Technology Company | Information Technology manager |
| Expert 6 | Tertiary | 30 | Technology Company | Software developer and analyst |
| Expert 7 | Tertiary | 38 | Technology Company | Systems Analyst |
| Expert 8 | Tertiary | 24 | Telecommunication company | Computer Research Scientist |
| Expert 9 | Tertiary | 33 | Technology Company | Data scientist |
| Decision-makers for BWM | | | | |
| DM1 | Tertiary | 32 | Technology Company | Manager |
| DM2 | Tertiary | 20 | Food processing company | Assistant Manager |
| DM3 | Tertiary | 22 | National Disaster Management Organisation | Safety manager |
| DM4 | Tertiary | 30 | Fire service | Battalion chief |
| DM5 | Tertiary | 18 | Ministry of Food and Agriculture (MOFA) | Safety manager |
| DM6 | Tertiary | 25 | Manufacturing -plastic | Assistant Manager |
| DM7 | Tertiary | 27 | Forestry Commission | Operations Manager |
| DM8 | Tertiary | 17 | Municipal Assembly | Director |
| DM9 | Tertiary | 21 | Technology Company | Assistant Manager |
| DM10 | Tertiary | 29 | Technology Company | Operations Manager |
| DM11 | Tertiary | 34 | Telecommunication company | Manager |
| DM12 | Tertiary | 25 | Forestry Commission | IT head |
| DM13 | Tertiary | 20 | Food processing company | Manager |
| DM14 | Tertiary | 23 | Ministry of Food and Agriculture | Project manager |
| DM15 | Tertiary | 25 | Fire service | Assistant chief |
| DM16 | Tertiary | 20 | Technology Company | Operations Manager |
| DM17 | Tertiary | 30 | Manufacturing -automobile | Production manager |
| DM18 | Tertiary | 15 | Food processing company | Manager |
| DM19 | Tertiary | 29 | Telecommunication company | Manager |
| DM20 | Tertiary | 16 | Manufacturing-textile | Assistant Manager |

All respondents had at least 15 years of experience. It took two months to finish the whole process through email and zoom meetings. The first question in the first round of the survey was posed as follows:

**Question 1:** " In your opinion, observation, and experience, what are the potential technologies to reduce the negative impact of extreme weather events (e.g., natural fire) on sustainable food and agriculture production? These could be current or perceived future technology based on technological trends and social change ."

**Question 2:** " In your opinion, opinion, observation, and experience, what are the roles of these technological platforms, challenges, and social factors? The social factors could be the environmental and social setting influencing the adoption of AI and supporting technologies. The roles could be the characteristics, capabilities or attributes of these technologies."

Through the first round of research, we collected about 8 technologies relevant to extreme weather context and a large number of their attributes in the second round, as shown in Figures 3 and 4. Experts used the 4-point scale (1 = not relevant, 2 = somewhat relevant, 3 = quite relevant, and 4 = highly relevant) to answer Question 1 and provided the relevant degree of these technologies. The 4-point scale is used to avoid a neutral midpoint.

Second, we use the content validity index (CVI) to confirm the valid attributes 1 and 2 indicate that the expert considers the attribute irrelevant to this study, and 3 and 4 indicate that the expert considers the attribute relevant to this study. Of the 9 experts, we think that 78.8% of the experts agree that this attribute is valid (Lynn, 1986). This means two experts could give 1 (not-relevant) or 2 (somewhat relevant) ratings, and this attribute will still be valid.

After the experts' responses, we computed the content validity of the individual technologies and proceeded to the evaluation of their attributes with BWM without a scale validity index. The 9 experts assessed the technologies platforms, as seen in Figure 7 B, and the CVI was used to validate the responses of the Delphi technique in the second round of the survey. Experts assessed and ranked the technologies required for extreme events and indicated their attributes and social factors that influence their full-scale adoption. These challenges are the drawbacks of technological implementation, while the social factors provide an atmosphere for implementation. About 100% of experts tipped ML as the most critical technology to reduce the impact of extreme events in the first phase of the survey; thus, subsequent rounds sought to explore the attributes, challenges, and social factors of implementing ML.

Third, experts categorize the attributes of AI into the four criteria of the UTAUT framework, which are performance expectancy(PE), effort expectancy(EE), social influence(SI), and facilitating conditions(FC) (Taherdoost, 2018), as seen in Table 2.

Best Worst Method is used to evaluate the relevance of these attributes to potential adoption grounds concerning extreme events. The roles and social conditions of AI were used to define the hierarchical framework to be assessed by decision-makers. Three social conditions were given by experts, and all of them were included in the BWM evaluation.

*Table 2  Attributes and validation based on UTAUT and expert opinion*

| Roles | E1 | E2 | E3 | E4 | E5 | E6 | E7 | E8 | E9 | Expert agreement | I-CVI | validation |
|---|---|---|---|---|---|---|---|---|---|---|---|---|
| Analytics of events | 4 | 3 | 4 | 3 | 4 | 4 | 3 | 3 | 4 | 9 | 1 | Valid |
| Predictive natures | 3 | 4 | 4 | 4 | 3 | 3 | 4 | 4 | 4 | 9 | 1 | Valid |
| Self-aware/self-reliant | 4 | 4 | 3 | 3 | 4 | 4 | 4 | 3 | 4 | 9 | 1 | Valid |
| Reduce human error and risk | 4 | 3 | 3 | 4 | 3 | 3 | 3 | 3 | 4 | 9 | 1 | Valid |
| Interoperability | 4 | 3 | 3 | 4 | 4 | 4 | 4 | 3 | 3 | 9 | 1.000 | Valid |
| Self-optimize | 4 | 3 | 3 | 2 | 4 | 3 | 4 | 3 | 4 | 8 | 0.889 | Valid |
| Better forecast | 4 | 3 | 3 | 3 | 4 | 3 | 4 | 3 | 4 | 9 | 1 | Valid |
| Availability of computing power | 3 | 4 | 3 | 4 | 4 | 3 | 3 | 3 | 3 | 9 | 1 | Valid |
| Increase productivity | 4 | 3 | 4 | 4 | 4 | 3 | 4 | 3 | 4 | 9 | 1 | Valid |
| Job provision | 4 | 3 | 3 | 4 | 4 | 3 | 4 | 3 | 4 | 9 | 1 | Valid |
| Human-machine interactions | 4 | 3 | 3 | 4 | 4 | 3 | 4 | 3 | 2 | 8 | 0.889 | Valid |
| Big data availability | 4 | 3 | 4 | 4 | 4 | 3 | 4 | 4 | 4 | 9 | 1 | Valid |
| Mass market potential | 2 | 3 | 3 | 4 | 4 | 3 | 4 | 3 | 4 | 8 | 0.889 | Valid |
| Facilitates decision making | 4 | 3 | 3 | 4 | 4 | 3 | 4 | 3 | 4 | 9 | 1 | Valid |
| Lifestyle enhancement | 2 | 3 | 3 | 3 | 3 | 4 | 3 | 4 | 3 | 8 | 0.889 | Valid |
| Increases automation | 4 | 3 | 3 | 4 | 4 | 3 | 4 | 3 | 4 | 9 | 1 | Valid |
| Personalization of task | 2 | 3 | 2 | 3 | 4 | 1 | 3 | 2 | 4 | 5 | 0.556 | Invalid |
| Enhances user experience | 4 | 2 | 3 | 4 | 3 | 2 | 3 | 2 | 4 | 6 | 0.667 | Invalid |
| - | - | - | - | - | - | - | - | - | - | - | - | - |
| - | - | - | - | - | - | - | - | - | - | - | - | - |
| - | - | - | - | - | - | - | - | - | - | - | - | - |
| - | - | - | - | - | - | - | - | - | - | - | - | - |
| | | | | | | | | | | S-CVI/Average | 0.932 | |
| | | | | | | | | | | Total agreement | 12.000 | |
| | | | | | | | | | | S-CVI/UA | 0.8 | |

## 3.2 Best Worst Method

The BWM, introduced by (Rezaei, 2015), is a simple and impactive multi-criteria decision-making (MCDM) method to determine attribute weights. BWM uses fewer pairwise comparisons than the analytic hierarchy process (AHP) method and improves the consistency and accuracy of judgment (Bai et al., 2020)(Safarzadeh et al., 2018)). The BWM includes five steps as follows (Rezaei, 2015).

Experts in the BWM differed from the experts in the Delphi survey, which is to avoid response bias. We used 20 decision-makers to evaluate the importance of attributes for AI as seen in Table 3 and 4.

**Step 1.** Let group decision-makers (DMs) construct a set of attributes $\alpha = \{c_j | j = 1,2,\cdots,n\}$ to be evaluated, where $n$ represents the number of attributes.

**Step 2.** Identify the best attribute $B$ and the worst attribute $W$ based on the opinions of group DMs.

**Step 3.** Create DMs' preference comparisons using the 9-point Likert scale for the best attribute compared to other attributes. The Likert scale ranges from 1 (equal importance between components) to 9 (much more significant than other attributes).

$$M_B = (m_{B1}, m_{B2}, \ldots m_{Bn})$$

**Step 4.** Similar to Step 3, create DMs' preference comparisons using the 9-point Likert scale for other attributes' compared to the worst attribute.

$$M_W = (m_{1W}, m_{2W}, \ldots m_{nW})^T$$

Table 3 Pairwise comparisons between attributes of AI extreme weather events

| Dimensions | No. of the Best Dimension | No. of the Worst Dimension | Attributes | codes | No. of the Best attribute | No. of the Worst attribute |
|---|---|---|---|---|---|---|
| Performance Expectancy (PE | 8 | 1 | Predictive nature | PE1 | 10 | 1 |
| | | | Events analytics | PE2 | 7 | 6 |
| | | | Better forecast | PE3 | 3 | 13 |

| Dimension | | | Attributes | | | |
|---|---|---|---|---|---|---|
| Effort expectancy (EE) | 2 | 9 | Ease of automation | EE1 | 4 | 10 |
| | | | Ease of learning | EE2 | 7 | 5 |
| | | | Ease of adoption | EE3 | 9 | 5 |
| Social influence (SI) | 7 | 0 | Political pressure | SI1 | 3 | 14 |
| | | | Technological giants' push | SI2 | 11 | 1 |
| | | | Social pressure | SI3 | 6 | 5 |
| Enabling conditions (EC) | 3 | 10 | Availability of computing power | EC1 | 6 | 1 |
| | | | Interoperability | EC2 | 2 | 1 |
| | | | Big data availability | EC3 | 12 | 0 |
| | | | Mass market potential | EC4 | 0 | 18 |

**Step 5.** Build the minimax model and solve optimal weights for all attributes. The weights of the attributes are obtained to minimize the highest absolute variations for all attributes, i.e. $\{|w_B - m_{Bj}w_j|, |w_j - m_{jw}w_w|\}$. Then, we can obtain the following minimax model:

$$\min \max \{|w_B - m_{Bj}w_j|, |w_j - m_{jw}w_w|\},$$

Subject to:

$$\sum_j w_j = 1;$$
$$w_j \geq 0, \forall j.$$

Convert the model into a linear programming model shown as:

$$\min \xi_L$$

Subject to:

$$|w_B - m_{Bj}w_j| \leq \xi_L^* \text{ for all } c_j$$
$$|w_j - m_{jw}w_w| \leq \xi_L^* \text{ for all } c_j$$
$$\sum w_j = 1;$$
$$w_j \geq 0, \forall j.$$

Table 3 Evaluation of AI-enabled extreme weather factors based on the UTAUT framework

| Dimension | Weight | Ranking | Attributes | Local weight | Local Ranking | Global weight | Global Ranking |
|---|---|---|---|---|---|---|---|
| Performance Expectancy (PE) | 0.339 | 2 | PE1 | 0.346 | 1 | 0.117 | 1 |
| | | | PE2 | 0.292 | 2 | 0.099 | 3 |
| | | | PE3 | 0.207 | 3 | 0.070 | 9 |
| Effort expectancy (EE) | 0.321 | 3 | EE1 | 0.173 | 3 | 0.059 | 11 |
| | | | EE2 | 0.226 | 2 | 0.072 | 7 |
| | | | EE3 | 0.234 | 1 | 0.075 | 6 |
| Social influence (SI) | 0.287 | 4 | SI1 | 0.106 | 3 | 0.034 | 12 |
| | | | SI2 | 0.331 | 1 | 0.095 | 4 |
| | | | SI3 | 0.238 | 2 | 0.068 | 10 |
| Enabling conditions (EC) | 0.342 | 1 | EC1 | 0.322 | 2 | 0.092 | 5 |
| | | | EC2 | 0.299 | 3 | 0.071 | 8 |
| | | | EC3 | 0.351 | 1 | 0.101 | 2 |
| | | | EC4 | 0.119 | 4 | 0.041 | 13 |

### 3.3 Machine Learning

The machine learning models considered in this work include linear regression, decision tree, random forests, and XGBoost. The descriptive statistics method guides and fast-track feature selection during extreme weather prediction. The dataset utilized for this is a publicly available research dataset (Cortez and Morais, 2007). This repository contains 517 instances and is reliable and valuable for research (Cortez and Morais, 2007), which is available at (https://archive.ics.uci.edu/ml/datasets/forest+fires). However, we contributed by employing a Tabular Generative Adversarial Networks (TGAN) to generate synthetic data to improve the models' prediction. Generative Adversarial Networks (GAN) have been used to generate an artificial dataset for research in several studies (Frid-Adar et al., 2018). Details of machine learning models used for predicting fires are outlined below;

3.3.1 Linear Regression (LR): LR is a supervised machine learning algorithm that executes a regression operation. The regression models the target variable based on the independent variable. It is used to find how variables relate to each other(Maulud & Abdulazeez, 2020).

3.3.2 Random forest (RF): A random forest comprises numerous independent decision trees that work together as an ensemble(Speiser et al., 2019). Each DT in RF spits a class of prediction, and the class with the highest votes becomes the model's prediction.

3.3.3 Support Vector Regression: SVR is a supervised learning approach used to forecast discrete values. SVR operates on the same basis as a support vector machine. The fundamental premise of SVR is to locate the best-fit line which has the maximum points called the hyperplane. These hyperplanes act as boundaries that inform decisions to predict the continuous output. See (Dash et al., 2021) for the mathematical formulation and more information about SVR.

3.3.4 Neural networks(NN): NN are also referred to as artificial neural networks (ANN), which are a subset of ML. The structure of NN is modeled after the human brain, mirroring the communication between organic neurons. An NN that consists of more than three layers is referred to as "deep learning." In the model, we used a basic neural network of two layers since the dataset was not very huge(Samek et al., 2021).

3.3.4: XGBoost regression (XGBR): An ensemble of weak prediction models serves as the output of extreme gradient boosting. For regression and classification issues, these models are typically decision trees. EGB creates a tree ensemble model from a collection of classifiers and RT's that aim to define and optimize an objective function by maximizing any differentiable loss function on a suitable cost function(Zhang & O'Donnell, 2020).

Detailed steps for data acquisition and training, testing, and prediction are outlined below.

**Step 1.** Data preparation for training
The original forest fire dataset contains 517 instances, a small sample size for training machine learning models. We used tabular generative adversarial networks to generate synthetic data to compensate for the smaller sample size. First, the original data was split into 0.6 test and 0.4 train sets before passing it into the TabGans network. This was to preserve some originality of the data in the test and validation phase.

**Step 2.** Generate synthetic data and standardization
A generative adversarial network was designed by Lan and colleagues in 2014(Goodfellow et al., 2014). With a given training set, Gans can generate a new data set with the same statistics as the original data. This technique preserves some degree of originality to the synthetic data. A Gan network is formulated as a game called GAN game between generator "G" and discriminator "D," as seen below:
For each probability space $(\delta, \mu_a)$ which sets the Gan game, where $\mu_a$ is the reference dataset.

The generator's strategic set is ($\rho(\delta)$) for all probability measures of $\mu_G$ on ($\delta$). The discriminator; strategy set follows a Markov kernels $\mu_D: \rho[0,1]$, $\rho[0,1]$ is the set of a probability measure on $[0,1]$. The GAN game follows the zero-sum game, which has an objective function as:

$$L(\mu_G, \mu_D) \coloneqq \mathbb{E}_{x\sim\mu_a, y\sim\mu_D(x)}[log y] + \mathbb{E}_{x\sim\mu_G, y\sim\mu_D(x)}[\log(1-y)]$$

The goal is for the generator to minimize the objective whiles the discriminator seeks to maximize the objective. The task of the generator is to approach $\mu_G \approx \mu_a$. This means the generator matches its output distribution as close as possible to the reference data. The version of GANs called Tabular GANs, which uses a similar game approach as above, was used to generate the synthetic to compensate for the smaller sample size. The generated data were concatenated to the train-set to train the machine learning models. After data generation, both the test and train test was standardized between 0 and 1, the standard scalar from the Sklearn preprocessing module, to minimize the effect of outliers and increase model performance.

**Step 3. Testing and evaluation**

Each of the machine learning was used to predict the x_test set and evaluated with the y_test and predicted_value. We used the Root mean Square and Mean Absolute Error to evaluate the performance of the machine learning models, as seen in Equations 1 and 2.

$$\text{RMSE} = \sqrt{\sum_{i=1}^{n}(x_{true,i} - x_{pred,i})^2} \tag{1}$$

$$\text{MAE} = \frac{\sum_{i=1}^{n} |x_{pred,i} - z_i|}{n} \tag{2}$$

Where $x_{true}$ is the actual burned area and $x_{pred}$ is the predicted burned area at the time i. $z_i$ is the observed burned area.

**Step 4.** Prediction

We used the best model(NN) to predict the fire burn area, which is closely related to the severity of the fire. Higher values mean more server fires and faster spreading fires, which is possible to cause more catastrophic effects and impede production, as seen in Figure 9. The feature importance influencing fires was determined using the Random forest feature importance technique, as shown in Figure 4

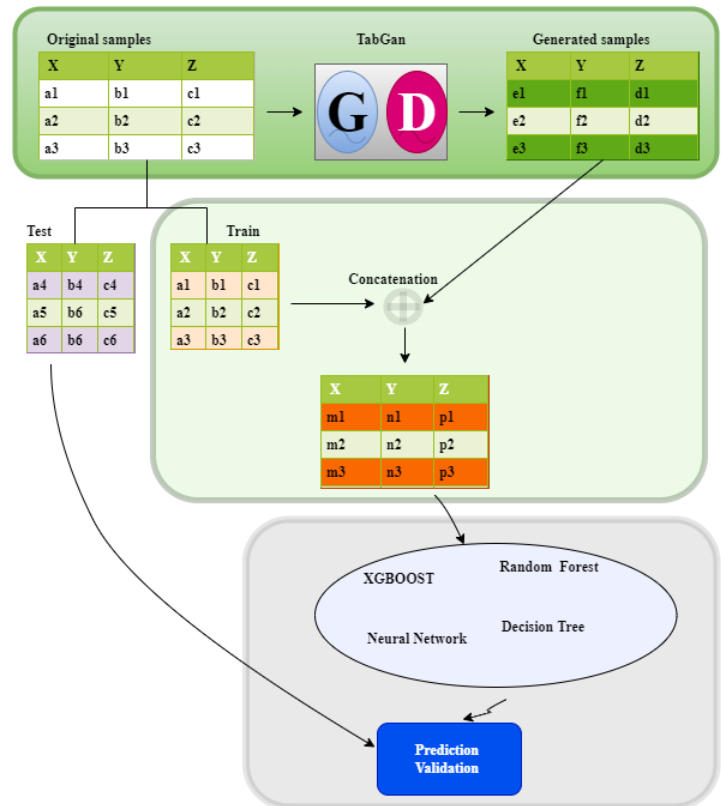

Figure 2 Systematic view of fire prediction model

### 3.5 Relationship Between the Methods Used
The relationship between these stand-alone methodologies used in this work is shown in Figure 5. The Delphi method was used to assess technologies relevant to reducing the impact of an extreme event and the roles and challenges of these technologies. Secondly, the 9 experts from the Delphi survey reviewed and categorized the attributes of AI into the UTAUT framework, which was later evaluated by 20 decision-makers. The relevance of the attribute of AI was assessed using the BWM. Afterward, the highly ranked technology from the Delphi method was used to affirm the most relevant attribute from the BWM using a publicly available forest fire dataset. The idea was not to complicate these straightforward methodologies but to comprehensively harness their potential in buttressing each other in reducing the impact of extreme weather on supply chain operations and production chains.

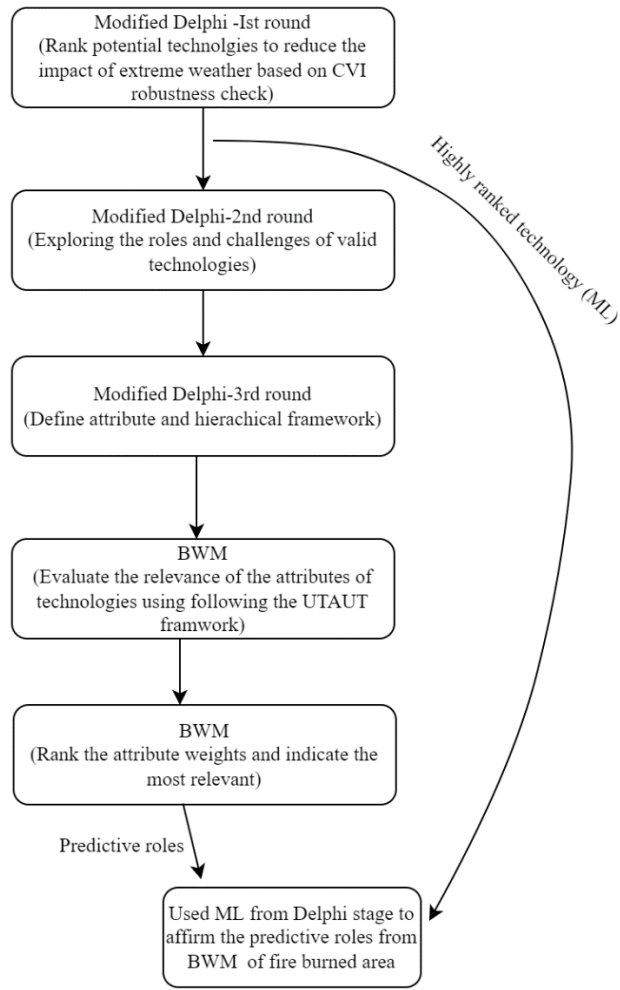

*Figure 3 Methodoligcal relationship*

## 4.0 Results and discussion

### 4.1 The dimension of attributes of AI for reducing extreme weather events

Among the 18 roles of AI using the expert opinion lens, 16 were valid in contributing to reducing extreme weather events. As seen in Table 2 above, these 16 roles were structured into 13 attributes and evaluated by BWM. Again, experts from the Delphi phase guided the categorization of these attributes, characteristics, and benefits into the UTAUT framework. The experts did not consider job provision and enhancement of user experience as required for reducing the adverse impacts of extreme events. Computing power and big data availability as characteristics of AI are considered hereafter as enabled technologies to boost AI functionalities. Ease of interoperability is a crucial attribute of AI, allowing it to connect to a broad technological ecosystem.

### 4.2 The important attributes of AI for reducing extreme weather events

In Figure 4 B, Artificial intelligence was ranked highest by all experts based on CVI robustness check, which indicates its potential to reduce extreme events. From table 5, the predictive nature of AI was highly ranked among decision-makers. The results showcase the concern and negative impacts of unpredictable extreme weather. Predictability is one of the major capabilities of AI and plays a critical role in reducing the impact of extreme events. (Jonkman et al., 2009) reported about 771 fatalities by hurricanes Katarina occurred in New Orleans. How would AI have helped? Deployment of autonomous robots on rescue missions reduces the time delay by human rescuers. For immediate medical care, fast and timely diagnoses by AI systems and robots in hospitals foster humanitarian disaster response.

Moreover, accurate predictions of extreme events enhance early warning systems for the populace. Likewise, during the covid-19 pandemic, extreme events disrupted the production of goods and services and caused a loss of infrastructure and a high economic downturn. Both big data availability and predictive nature were ranked first locally, but the latter was second to the fore globally. Big Data and computing power aid AI techniques in accurately predicting extreme events to hike early warning criteria. Ease of learning drives the self-optimization and reliance of AI technologies on a new set of data and trends. Auspiciously, the availability of computing power and big data provides the platform for advancing AI. As mentioned in section 2, data quality and energy use are crucial challenges to AI transition. Hence there should be a keen interest in designing climate-smart AI infrastructure to reduce the downside of the technology and the impact of the extreme event and beyond.

Technological giant's within the ecosystem of emerging technologies are critical factors of social concern. Will there be a power shift and control to highly capable tech companies? The future of such debates is uncertain, and the study heeds interest in the social dimension of using AI and supporting technologies even though the social influence was ranked least among the four dimensions. Ease of automation, ease of adoption, and interoperability had almost similar global weightings. The interoperability of AI with other technologies is a key enabling factor in the digital ecosystem. Particular autonomous robots and wide-scale adoption in the manufacturing industry. The ease of adoption of autonomous robots in firefighting and rescue missions is imperative to reducing the impacts of extreme events. Political pressure in adopting AI and digital technologies was ranked last but not least, which means the political dimension was not considered a top influencer in adopting AI for extreme weather or social pressures (ranked 10$^{th}$). The focus is more on the capabilities and enabling environment for achieving the full potential of AI. However, we should not disregard the impact of government policies in adopting digital technologies in general. The market potential of AI is a critical factor in its wide-scale adoption, but it was last ranked in the context of extreme weather.

Nevertheless, there were challenges to the adoption of these technologies; key among them was diversity in data systems, job displacement, sourcing of talents, machine-to-machine variations, power balance and ownership, security and privacy, and environmental impacts (Narwane et al., 2022) as seen in Figure 4 A. (Narwane et al., 2022), (Bibi et al., 2017) and (Sinha et al., 2019) analyze similar challenges in the adoption of IoT in agriculture and food supply chains which affirms the finding of this research.

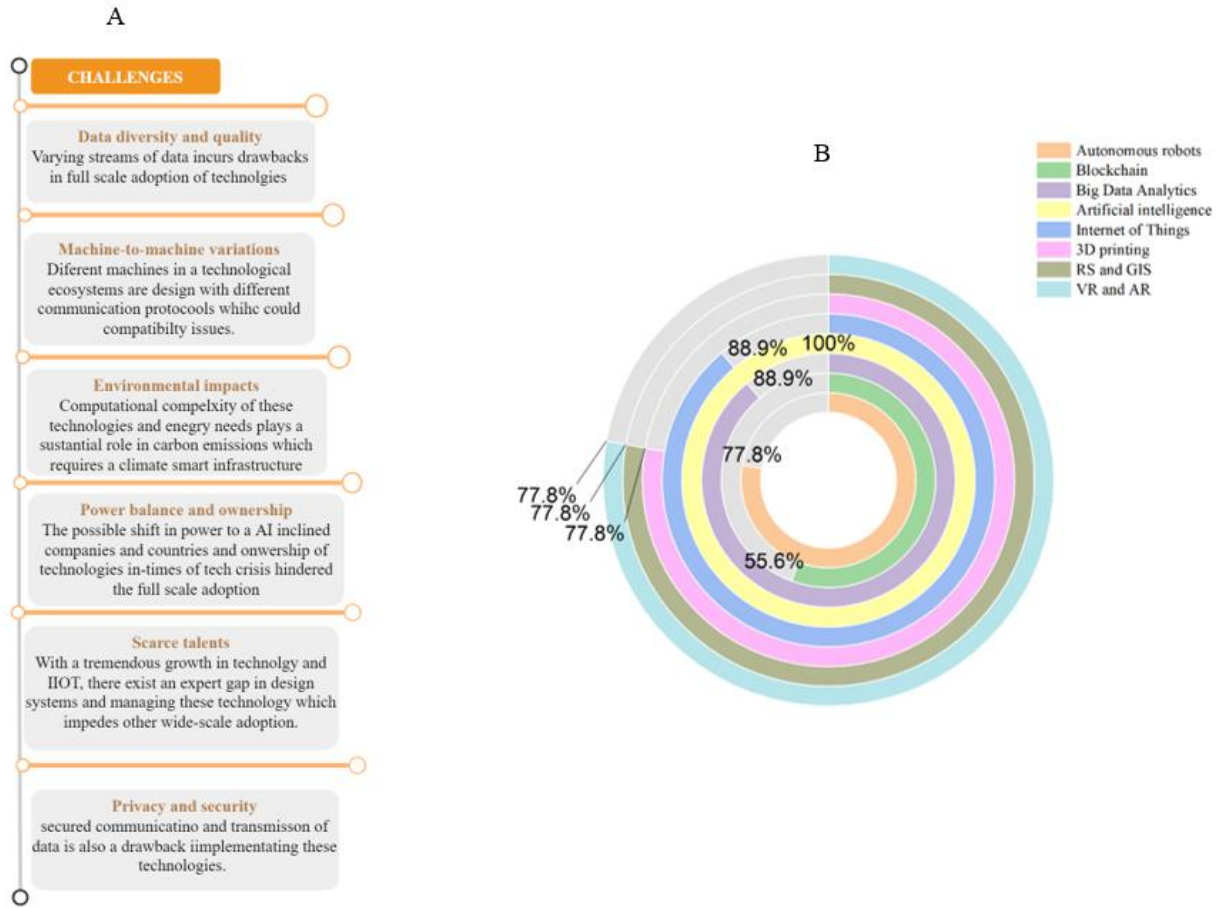

*Figure 4 Challenges of adopting technologies (A) and Validation of technolgies for reducing the impact of extreme events (B)*

**4.1 Machine learning**

The performance of the machine learning models is shown in the Table 5. For the two metrics considered, neural networks were the best-performing model compared to all the machine learning models with an RMSE of 0.0093 and MAE = 0.0717. Random Forest closely matched the NN in prediction fires. Each randomized tree in the RF model is constructed from a sample taken with replacement from the training set using the specified parameters. The averaging of the independent and adjusted trees produced a superior prediction model overall, and this experiment proved those improvements. However, the NN, which is inspired by biological neural networks, performed better than all the other models. The original data had 517 instances which was a smaller sample size for training ML models. The results in Table 6 buttress the need for a huge dataset for better performance of ML models (Sharma et al., 2021). For all the meteorological and FWI parameters for fire prediction, the rain was very influential in fire burn area prediction. The combined risk of all these predicting parameters indicates the possible occurrence of fire or not. Both temperatures and rain are determinants of the fine fuel moisture code (FFMC) and Duff moisture code (DMC). A region with less rainfall and extreme and dry winds become a high fire-risk area. As seen in Figure 7, fluctuations in rainfall and temperature are high determinants of fires since a dried fuel ignites easily and spreads faster on a windy day. Fire prevention strategies and interventions should be stringent in less rainfall, high temperatures, and windy areas. These three factors influence drought, DMC, and FFMC in forest areas. Farmers should be prevented from preparing their lands with fires or highly guided with prescribed burning strategies during periods of fewer spots of rain, especially drought. Arsons should be severely punished to deter people from burning public properties.

Table 4 Models for Predicting Fires

| Models Considered | With TabGAN (contribution) | | Without TabGAN | |
|---|---|---|---|---|
| | RMSE | MAE | RMSE | MAE |
| NN (200 epochs) | 0.0093 | 0.0717 | 2.1002 | 0.6429 |
| NN (100 epochs) | 0.0355 | 0.1297 | 1.2415 | 0.5609 |
| Linear Regression | 0.3879 | 0.3554 | 0.71593 | 0.3738 |
| Random Forest | 0.0225 | 0.05434 | 1.3568 | 0.4093 |
| XGBRegressor | 0.0796 | 0.14917 | 1.4306 | 0.3696 |
| SVR | 0.3062 | 0.15152 | 0.6945 | 0.2091 |

Figure 5 indicates the prediction of the fire-burned area compared to the actual values. The difference between the predicted and real values were not statistically significant (P-value <0.05 at a 95% confidence interval). This means that the apparent difference between the two happened by chance. Thus, the model could predict fire burn area given all meteorological and FWI parameters. These predictions buttressed early warning schemes to guide businesses and farmers to make informed supply chains and farm decisions. (Despoudi et al., 2021) concluded contingency theory that supply chain management is more unpredictable when there is a large degree of environmental disturbance. Likewise, with the COVID-19 pandemic, the impact of extreme events is problematic to sustainable supply chains, particularly food supply chains. Predictive technologies such as Machine learning are imperative to forecast the occurrence of extreme events to boost supply chain resilience. In this current work, we utilized machine learning models to predict fire-burned areas of a publicly available forest fire dataset in figure 5 and the contributing factors to fires in Figure 6. Auspiciously, ML can also forecast supply chain inventory, demand and supply, and lead time, which is indispensable for supply chain resilience to extreme events and resources.

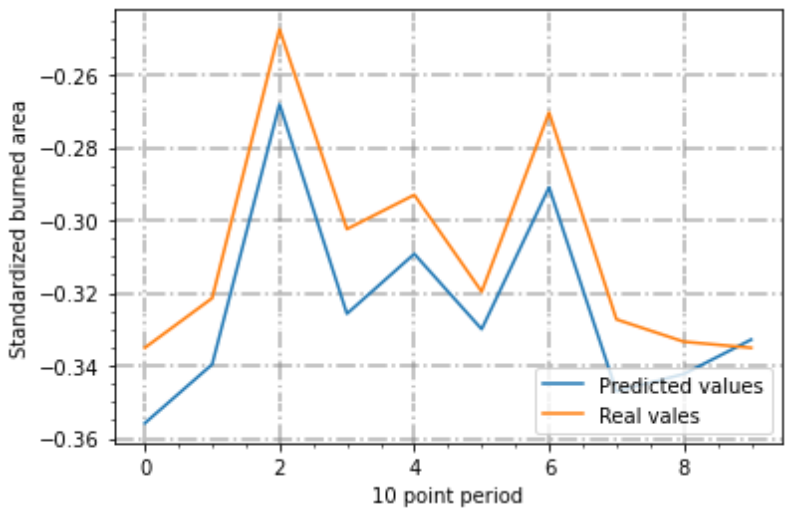

Figure 5 Burned area prediction using ML

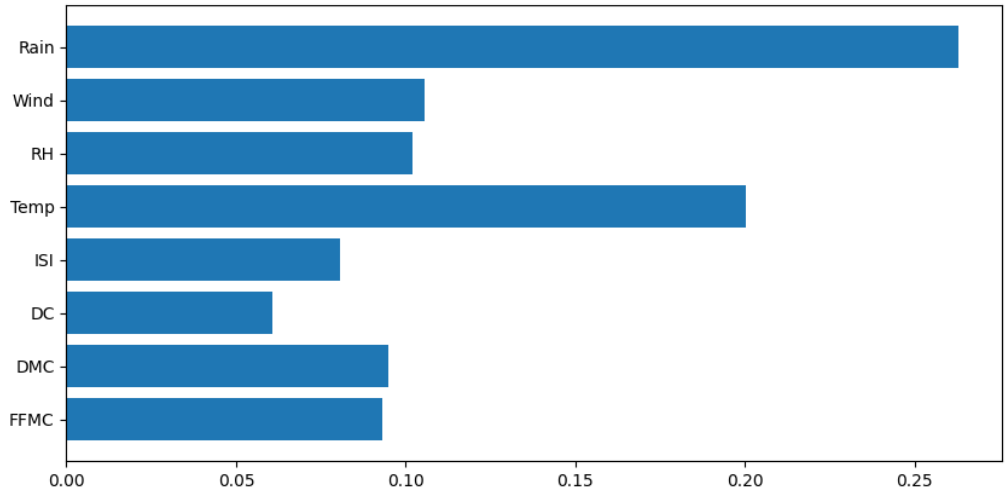

Figure 6 Weights of Features Driving Forest Fires

## 5.0 Implications
### 5.1 Managerial implications

Enterprises, technology managers, and inventors will find this work's findings and methodology more significant to their organizations.

First, the attribute framework in Table 3 will guide firms in making informed decisions during entrepreneurial orientation and environmental dynamism scenarios. The roles outlined in Table 2 vividly clarify the potential of AI and supporting technologies that firms could infer when adopting technologies that can comprehensively provide these roles. Moreover, the validation of technologies in Figure 4A provides the technological set that firms could adopt to boost their operations in pre- and post-extreme events.

Second, the weights of attributes in table 2 posit the level of importance of these attributes. These degrees of relevance apprise firms in allocating scarce resources and investments to meet the most pertinent attribute and select the set of technologies that could provide these attributes. Moreover, the enabling condition of these technologies is significant for achieving the intended operational performance of these technologies.

Third, the prediction potential of machine learning, as tested in figure 5, indicates AI capabilities that firms could adopt. Firms could align their organizational capabilities to suit the implementation of predictive systems. Unpredictable events have proven catastrophic to supply chains, and lessons from the Covid-19 pandemic and observing societal disruption affirms the chaos of extreme events. Predictive systems for reducing extreme events will have a plus for firms. These pluses come about when predicting lead time, inventory, and stocks and forecasting supply and demand. Artificial intelligence and supporting technologies cede the platform for firms to design better resilient strategies to cushion supply chain and operational shocks. Technological readiness is one of the key critical success factors of adopting AI in food supply chains (Dora et al., 2022). Our findings highlight supporting technologies and enabling the environment to improve supply chain resilience and coordination.

Finally, environmental dynamism shows many uncertainties in long-term decision-making, particularly concerning technological inclusion in operational management and supply chain performance. To guide comprehensive and strategic planning, using mixed methods to explore varying dimensions of technologies, trends, degree of importance, and key capabilities is critical to firms' performance.

## 5.2 Methodological Implications

The proposed Delphi BWM method in developing attributes is essential in evaluating AI's role and capabilities. We utilized a qualitative analysis and predictive technique to prove the potential of AI in reducing the impact of extreme events.

First, the use of AI in the business environment has suffered from reaching its full performance gains(Fountaine et al., 2019) (Ransbotham et al., 2018). (Brynjolfsson et al., 2018) terms this as a modern productive paradox. The authors affirm that implementation and reorganization delays are key reasons AI has not produced desired results. In order to realize performance improvements from AI, it is crucial to comprehend the complementary resources that need to be developed and put them into practice. Thus, multi-criteria decision-making and expert opinion ascertain the nitty-gritty of AI to highlight the attributes, roles, and requirements of AI to attain their imaginable results. The predictive capability of AI, as shown in this research, is highly dependent on large and quality data and higher computing power confirming the findings from the BWM.

Second, firms could follow the methodologies utilized in this work and findings to posit the attributes of AI and supporting technologies in predicting and forecasting extreme events to advance early warning systems. Such early warning systems contribute to the reduction of extreme weather.

The extensive methodology methodologies utilized in this work could be applied in related fields such as technology forecasting and evaluation of technological capabilities to boost supply chain resilience in extreme events.

## 6.0 Conclusion

In this paper, we explored the possibility and avenues of AI and supporting technologies in reducing the impact of extreme weather events. The Delphi method was used to define and validate the attributes of AI in reducing extreme events. Following the UTAUT framework, the BWM methodology was used to evaluate valid attributes of AI. It was revealed that facilitating conditions are highly relevant in AI applications to extreme events, with predictive roles as the most prominent. Technology is fast advancing with social change. It is critical to assess the potential of technologies in meeting societal transitions to improve and design suitable technologies to meet societal demands. Our proposed hierarchical framework gives a precise scenario of the roles of AI and supporting technologies in reducing the impact of extreme weather. We presented machine learning techniques for predicting forest burned areas using a publicly available dataset from Portugal. The insignificant difference between the predicted and real values confirms the higher predictive ability of machine learning models.

Exploring AI and supporting technologies for climate-resilient production under extreme events define the research context. Even though there is the diverse theoretical and practical implications of this study, some limitations give rise to future works. The study focused on experts from the Ghanaian setting. This could breed jurisdictional bias in assessing AI's push factors and roles in reducing the impact of extreme events. These factors presented in the hierarchical framework may not be exhaustive since the goal was to obtain comprehensive factors that define AI. Based on the geographic location, expertise, and level of knowledge or experience, several factors could be added to evaluate the full potential of AI. Moreover, cause and effect relations could exist among these factors, which were not considered in this current research; thus, a multi-criteria decision-making approach such as DEMATEL could be harnessed in future works to elaborate on these relationships, particularly a research focus on AI and supporting technologies. Research on the roles of digital technologies in reducing the impact of extreme weather events is emerging, and there exist vast avenues in utilizing decision-making tools to evaluate the capabilities and characteristics of these technologies. We anticipate varying challenges to fully adopting these technologies. We believe there are overt challenges depending on these challenges and the use cases of digital technologies, such as extreme weather events, production and operation, and supply chain resilience. This work contributes to strengthening the research base on the roles of AI and supporting technologies in reducing the impact of extreme events.